# Opportunistic Multiuser Scheduling in Three State Markov-modeled Downlink

Sugumar Murugesan, Philip Schniter

*The Ohio State University*
*Department of Electrical and Computer Engineering*
*2015 Neil Avenue, Columbus, Ohio 43210*

August 29, 2008

**Abstract**

We consider the downlink of a cellular system and address the problem of multiuser scheduling with partial channel information. In our setting, the channel of each user is modeled by a three-state Markov chain. The scheduler indirectly estimates the channel via accumulated Automatic Repeat Request (ARQ) feedback from the scheduled users and uses this information in future scheduling decisions. Using a Partially Observable Markov Decision Process (POMDP), we formulate a throughput maximization problem that is an extension of our previous work where the channels were modeled using two states. We recall the greedy policy that was shown to be optimal and easy to implement in the two state case and study the implementation structure of the greedy policy in the considered downlink. We classify the system into two types based on the channel statistics and obtain round robin structures for the greedy policy for each system type. We obtain performance bounds for the downlink system using these structures and study the conditions under which the greedy policy is optimal.

## 1 Introduction

Opportunistic multiuser scheduling, introduced by Knopp and Humblet in [1] and defined as *allocating the resources to the user experiencing the most favorable channel conditions* has gained immense popularity among network designers in the recent past. Opportunistic multiuser scheduling essentially taps the multiuser diversity in the system and has motivated several researchers ([2, 3, 4, 5, 6]) to study the performance gains obtained by opportunistic scheduling under various scenarios. While i.i.d flat fading model is popularly used by researchers in modeling time varying channels, it fails to capture the memory in the channel observed in realistic scenarios. This motivated the researchers to consider the Gilbert Elliott model [7] that represents the channel by a two state Markov chain. Specifically, a user experiences error-free transmission when it observes a "good" channel, and unsuccessful transmission in a "bad" channel. Several works have been done on opportunistic multiuser scheduling in this Markov modeled channel [8, 9, 10, 11, 12]. It is understandable that the availability of the channel state information at the scheduler is crucial for the success of the opportunistic scheduling schemes. Traditionally, when the scheduler has no channel information, pilot based channel estimation is performed and the estimates are used for scheduling decisions ([2, 6, 13]). A new line of work, [14, 15, 16, 17, 18], attempts to exploit Automatic Repeat reQuest (ARQ) feedback, traditionally used for error control at the data link layer, to estimate the state of the two state Markov modeled channels.

In [16] and [18], for a two state Markov modeled downlink (one to many communication) system, a greedy policy has been shown to be optimal from a sum throughput point of view. This greedy policy is shown to be amenable to an easy implementation with a simple round robin based algorithm that takes as input the ARQ feedback from the scheduled user. Although modeling the channel by a two state Markov chain is a welcome change from the traditional memoryless models, the scheduler can do better by discriminating the channel conditions on a finer level, i.e., if the channel is modeled by higher state Markov chains. As a first step in this direction, in this report, we model the channels by three state Markov chains and study the property of the greedy policy and conditions under which it will be optimal.

The report is organized as follows. The problem setup is described in Section 2 followed by a study of the implementation structure of the greedy policy in Section 3. A comparison of the original system with the genie-aided system, introduced in [18], will be made in Section 4. In Section 5, upper and lower bounds to the system performance is derived. We study the conditions under which greedy policy is optimal in Section 6. Conclusions are provided in Section 7.

## 2 Problem Setup

### 2.1 Channel Model - Probability Transition Matrix

We consider downlink transmissions with 2 users. The channel between the base station and each user is modeled by an i.i.d, first order, three-state Markov chain. Time is slotted and the channel of each user remains fixed for a slot and evolves into another state in the next slot according to the Markov chain statistics. The time slots of all users are synchronized. The three-state Markov channel is characterized by a $3 \times 3$ probability transition matrix

$$P = \begin{bmatrix} p_{11} & p_{12} & p_{13} \\ p_{21} & p_{22} & p_{23} \\ p_{31} & p_{32} & p_{33} \end{bmatrix}, \qquad (1)$$

where $p_{ij}$ is the probability of evolving from state $i$ to state $j$ in the next slot.

The states are made to represent the quantized strength of the channel, with state 1 assumed to represent the lower end of the channel strength spectrum and state 3 representing the higher end. We assume that the Markov chain is positively correlated in time. Thus $p_{ii} \geq p_{ji}$ if $j \neq i$. Also, motivated by observation of realistic channels, we assume that the channel evolves in a smooth fashion across time. Thus $p_{21} \geq p_{31}$ and $p_{23} \geq p_{13}$. Also, observing that state 3 represents a region of the channel strength spectrum that is not bounded from above, it is reasonable to assume $p_{32} \leq p_{12}$. To summarize, the transition matrix elements are related as below:

$$p_{11} \geq p_{21} \geq p_{31}$$
$$p_{22} \geq p_{12} \geq p_{32}$$
$$p_{33} \geq p_{23} \geq p_{13} \qquad (2)$$

## 2.2 Existence of Steady State

Let $\mathbf{p}_{ss}$ denote the steady state probability vector of the Markov chain with $\mathbf{p}_{ss}(i)$ representing the steady state probability of state $i$. We now rule out the instances of $P$ matrix entries that either 1) obviously lead to a steady state $\mathbf{p}_{ss}(i) = 0$ for some $i \in \{1, 2, 3\}$ or 2) eliminates the possibility of a steady state altogether. Both these cases trivialize the scheduling problem we address in this report. From the inequalities governing the elements of the $P$ matrix, we see that $p_{ii} > 0$. Otherwise, $p_{ji} = 0$ leading to $\mathbf{p}_{ss}(i) = 0$. Also $p_{12} \neq 0$. Since, if $p_{12} = 0$, then $p_{32} = 0$. Thus $p_{ss}(2) = 0$ (since when the channel enters state 1 or 3 it will never reach state 2 again). For similar reasons, $p_{21} \neq 0$ and $p_{23} \neq 0$. Thus the only elements that can be zero are $p_{13}$, $p_{31}$ and $p_{32}$. Among these $p_{31}$ and $p_{32}$ cannot be both zero. Otherwise, $p_{33} = 1$ making state 3 an absorbing state leading to $\mathbf{p}_{ss}(1) = \mathbf{p}_{ss}(2) = 0$. Thus there can be at most one zero in any row and at most one zero in any column of $P$. This, along with the fact that all the entries are non-negative, renders $P^2$ a positive matrix (i.e., all the elements are positive).

From [19] (p.51), *A nonnegative square matrix, $\mathbf{A}$ is said to be regular iff there exists a natural number $r$ such that $\mathbf{A}^r$ is a positive matrix.* Thus $P$ is a regular matrix with $r \geq 2$.

We now reproduce Theorem 4.2 from [19]

**Theorem 1.** *If $\mathbf{A}$ is a regular stochastic matrix then $\mathbf{A}^n$ converges as $n \to \infty$ to a positive stable stochastic matrix $e\Pi'$, where $\Pi = (\pi(i))_{i \in state\ space}$ is a probability vector with non null entries and $e$ is a unit vector with dimension equal to the cardinality of the state space.*

Thus the $n$-step transition probability matrices of the Markov channels in our problem also converge to stable stochastic matrices. Since this is necessary and sufficient for the existence of steady state, under the conditions established earlier, the Markov channels in our problem have steady state with the steady state probability vector given by $\mathbf{p}_{ss}$.

## 2.3 Scheduling Problem

The base station is the central controller that controls the transmission to the users in each slot. In any time slot, the base station does not know the *exact* channel state of the users and it must schedule the transmission of the head of line packet of exactly one user (a data queue is maintained for each user to collect the data meant for that user). Thus, a TDMA styled scheduling is performed here. The power spent in each transmission is fixed. At the beginning of each time slot, the head of the line packet of the scheduled user is transmitted and is dropped from the queue. The scheduled user, based on measurements of the signal strength of the received data packet, obtains information on the state of the channel and sends this back to the scheduler. We call this feedback as $F_i$ with $i \in \{1, 2, 3\}$. This channel state feedback is assumed to be transmitted over a dedicated error-free channel. This feedback information, along with the label of the slot in which it is acquired, will be used in future scheduling decisions. The performance metric that the base station aims to maximize is the sum reward of the system. Details will be discussed in the next section.

## 2.4 Formal Problem Definition

Since the base station must make scheduling decisions based on only a partial observation of the underlying Markov chain, we employ techniques from partially observable Markov decision process (POMDP) [20, 21, 22, 23] theory in this work. We now proceed to introduce the terms/entities that we use in this work, many of which are borrowed from the POMDP literature.

*Control interval $k$*: Each time slot in our problem setup will henceforth be called a control interval. The "end" of the POMDP is fixed. A control interval is indexed by $k$ if there are $k - 1$ more intervals until the end of the process.

*Action $a_k$*: Indicates the index of the user (1 or 2) scheduled in control interval $k$.

*Belief vector of user $i$ at the $k^{th}$ control interval $\pi_{k,i}$*: Element $\pi_{k,i}(j)$ denotes the probability that the channel of user $i \in \{1, 2\}$ in the $k^{th}$ control interval is in state $j \in \{1, 2, 3\}$, given all the past information about that channel. If $F_j$ was received from user $i$, $l + 1$ control intervals earlier with $l \in 0, 1, 2, \ldots$, then the belief vector in the current interval $k$ is given by $\pi_{k,i} = [p_{j1}\ p_{j2}\ p_{j3}]P^l$. We will henceforth represent the vector $[p_{j1}\ p_{j2}\ p_{j3}]$ by $\mathbf{p}_j$. If user $i$ is not scheduled in control interval $k$, then the belief vector of this user evolves to the next interval as follows: $\pi_{k-1,i} = \pi_{k,i}P$.

It has been proven in [20] that the belief vector $\pi_{k,i}$ is a sufficient statistic to any information about the channel of user $i$ in the past control interval, in our case, the scheduling decisions and the channel feedback information from the past. Thus the scheduling decision in any control interval can be solely based on the belief vectors for that interval and not on the past channel feedback or schedule information.

*Scheduling Policy $\mathfrak{A}_k$*: A scheduling policy $\mathfrak{A}_k$ in the control interval $k$ is a mapping from the belief vectors and the control interval index to an action as follows:

$$\mathfrak{A}_k : (\pi_{k,1}, \pi_{k,2}, k) \to a_k \quad \forall k \geq 1$$

Note that the scheduling policy can, in general, be time-variant.

*Reward Structure*: In any control interval $k$, a reward of $\alpha_i$ is accrued when the scheduled user sends back $F_i$. Let state 1 be defined such that no reward is accrued when an user in state 1 is scheduled, i.e., $\alpha_1 = 0$. This assumption can be satisfied by letting state 1 represent the channel strengths that do not allow any useful data transfer. Since state 3 represents channel strengths that are better than those represented by state 2, we have $\alpha_3 \geq \alpha_2$. Throughout this report we will assume $\alpha_3 = 1$ without loss of generality.

*Net Expected Reward in the control interval m, $V_m$*: With the belief vectors, $\pi_{m,1}$, $\pi_{m,2}$ and the scheduling policy, $\{\mathfrak{A}_k\}_{k \leq m}$, fixed, the net expected reward, $V_m$, is the sum of the reward, $R_m(\pi_{m,a_m}, a_m)$, expected in the current control interval $m$ and $\mathrm{E}[V_{m-1}]$, the net reward expected in the future control intervals conditioned on the belief vectors and the scheduling decision in the current control interval. Formally,

$$V_m(\pi_{m,1}, \pi_{m,2}, \{\mathfrak{A}_k\}_{k \leq m}) = R_m(\pi_{m,a_m}, a_m) \\ + \mathrm{E}[V_{m-1}(\pi_{m-1,1}, \pi_{m-1,2}, \{\mathfrak{A}_k\}_{k \leq m-1}) | \pi_{m,1}, \pi_{m,1}, a_m],$$

where the expectation is over the belief vectors $\pi_{m-1,1}, \pi_{m-1,2}$. With [1] $\alpha = [\alpha_1\ \alpha_2\ \alpha_3]^T$, the expected current reward can be written as

$$R_m(\pi_{m,a_m}, a_m) = \pi_{m,a_m}\alpha.$$

Note that if $a_m$ was observed to be in state $i$ in the previous interval then $\pi_{m,a_m} = \mathbf{p}_i$ and $R_m(\pi_{m,a_m}, a_m) = \mathbf{p}_i\alpha$.

*Performance Metric- the Sum Reward, $\eta_{sum}$*: For a given scheduling policy, $\{\mathfrak{A}_k\}_{k \geq 1}$, the sum reward is given by

$$\eta_{\mathrm{sum}}(\{\mathfrak{A}_k\}_{k \geq 1}) = \lim_{m \to \infty} \frac{V_m(\mathbf{p}_{ss}, \mathbf{p}_{ss}, \{\mathfrak{A}_k\}_{k \geq 1})}{m}, \tag{3}$$

where $\mathbf{p}_{ss}$ is the steady state probability vector of the underlying Markov channels.

*Optimal Scheduling Policy, $\{\mathfrak{A}_k^*\}_{k \geq 1}$*:

$$\{\mathfrak{A}_k^*\}_{k \geq 1} = \arg\max_{\{\mathfrak{A}_k\}_{k \geq 1}} \eta_{\mathrm{sum}}(\{\mathfrak{A}_k\}_{k \geq 1}). \tag{4}$$

## 3 Structure of the Greedy Policy

Consider the following policy:

$$\widehat{\mathfrak{A}}_k : (\pi_{k,1}, \pi_{k,2}, k) \to a_k = \arg\max_{a_k} R_k(\pi_{k,a_k}, a_k) \\ = \arg\max_i \pi_{k,i}\alpha \quad \forall k \geq 1.$$

Since the above given policy attempts to maximize the expected current reward, without any regard to the expected future reward, it follows an approach that is fundamentally *greedy* in nature. For this reason, we henceforth call $\{\widehat{\mathfrak{A}}_k\}_{k \geq 1}$ the Greedy Policy. We now proceed to derive the implementation structure of the greedy policy.

**Lemma 2.** *For any $k \geq 0$, the immediate reward expected by scheduling an user that was observed $k + 1$ control intervals earlier, to be in state 2, lies between the rewards corresponding to states 3 and 1, i.e,*

$$\mathbf{p}_1 P^k \alpha \leq \mathbf{p}_2 P^k \alpha \leq \mathbf{p}_3 P^k \alpha, \forall k \in 0, 1, 2, \ldots \tag{5}$$

---
[1] $\mathbf{x}^T$ indicates the transpose of vector $\mathbf{x}$.

**Lemma 3.** *The immediate reward expected by scheduling an user that was observed, $k+1$ control intervals earlier, to be in state 3, monotonically decreases to $\mathbf{p}_{ss}\alpha$ as $k$ increases from $0 \to \infty$, i.e.,*

$$\mathbf{p}_3 P^{k+1}\alpha \leq \mathbf{p}_3 P^k \alpha, \ \forall k \in 0, 1, 2, \ldots$$
$$\mathbf{p}_3 \lim_{k \to \infty} P^k \alpha = \mathbf{p}_{ss}\alpha \qquad (6)$$

Note that $\mathbf{p}_{ss}\alpha$ is the immediate reward expected when no past information about the user is available or when the belief vector of the user equals the steady state vector, $\mathbf{p}_{ss}$.

**Lemma 4.** *The immediate reward expected by scheduling an user that was observed, $k+1$ control intervals earlier, to be in state 1, monotonically increases to $\mathbf{p}_{ss}\alpha$ as $k$ increases from $0 \to \infty$, i.e.,*

$$\mathbf{p}_1 P^{k+1}\alpha \geq \mathbf{p}_1 P^k \alpha, \ \forall k \in 0, 1, 2, \ldots$$
$$\mathbf{p}_1 \lim_{k \to \infty} P^k \alpha = \mathbf{p}_{ss}\alpha \qquad (7)$$

Note that, from the above lemmas, we have

$$\mathbf{p}_2 \lim_{k \to \infty} P^k \alpha = \mathbf{p}_{ss}\alpha. \qquad (8)$$

In all the above results, the immediate reward approaches $\mathbf{p}_{ss}\alpha$ as the time since the last observation of the user increases. This is because, in the underlying first order Markov chain, the dependency between the states in two control intervals (memory) diminishes as the time gap between the control intervals increases. These lemmas are instrumental in obtaining the algorithm for implementing the greedy policy, that will be summarized soon. We first identify two types of system based on the property of the $P$ matrix and the reward values.

- Type I system: when $\mathbf{p}_2\alpha \geq \mathbf{p}_{ss}\alpha$
- Type II system: when $\mathbf{p}_2\alpha < \mathbf{p}_{ss}\alpha$

The implementation algorithm for the greedy policy significantly changes depending on the type of the system.

**Proposition 5.** *When the system is type I, the greedy policy is implemented as follows*

- *If feedback $F_3$ or $F_2$ was received from the user scheduled in the previous control interval (identified as user s), reschedule the user in the current control interval.*
- *Schedule the other user (identified as user u) if feedback $F_1$ was received.*

*Proof.* Refering to Fig. 1, when $F_3$ was received from user $s$, the expected reward if $s$ is scheduled again is given by $p_3\alpha$. The expected reward if $u$ is scheduled is a point on one of the three curves (for $k > 0$) in the figure. Note that $p_3\alpha$ is greater than any point (the y-dimension) on any of the curves, thus establishing 'retain the schedule if $F_3$ is received' policy. This result essentially stems from the following facts: 1) Higher reward ($\alpha_3 = 1$) is accrued when the scheduled user happens to be in state 3 than in other states. 2) The Markov channel is positively correlated in time ($p_{ii} \geq p_{ji}$ if $i \neq j$).

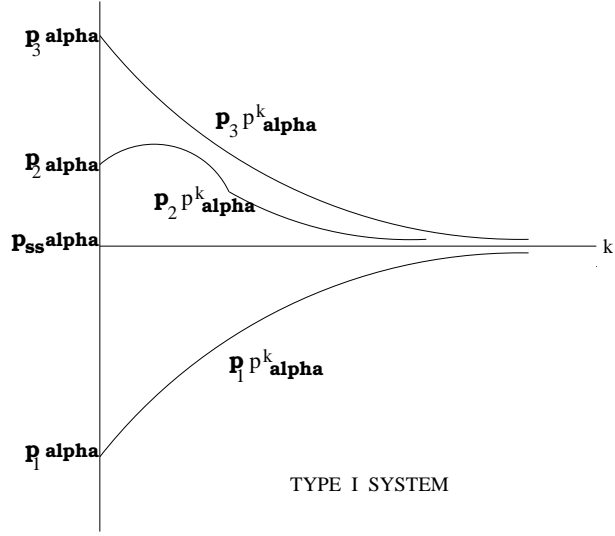

Figure 1: Type I system

Similarly when $F_1$ was received from user $s$, the expected reward if $s$ is scheduled again is given by $p_1\alpha$ which is less than any other point on the three curves, thus establishing 'switch if $F_1$ is received' policy.

When $F_2$ is received, assuming the greedy policy was implemented so far since the beginning of the scheduling process, the reward expected if $u$ is scheduled lies on the lower curve $\mathbf{p}_1 P^k \alpha$ for $k > 0$. This is because the first time (since the beginning of the scheduling process) a $F_2$ is received (call this interval $m_0$), if greedy policy was implemented so far, user $u$ (the waiting user) would not have given $F_3$ when it was dropped and since this is the first time $F_2$ is observed by the scheduler, $u$ would not have sent $F_2$ either, when it was dropped. Therefore $u$ must have sent $F_1$ the last time it was scheduled (and hence dropped). Thus the reward expected if $u$ is scheduled now (at $m_0$) falls on the bottom curve leading to retaining of user $s$ (since $\mathbf{p}_2 \alpha \geq \mathbf{p}_1 P^k \alpha$ for any $k \geq 0$). In the next instance of $F_2$ reception, the same logic holds (as long as greedy policy is implemented all along until this instance) and so on for subsequent instances of $F_2$. Note that the condition *greedy must be implemented since the beginning until 'now'* is quite natural given our interest in implementing the policy in the current interval. Thus there is no loss of generality here.

These arguments establish the proposition. □

**Proposition 6.** *When the system is type II, the greedy policy is implemented as follows*

- *If feedback $F_3$ was received from the user scheduled in the previous control interval (call it user $s$), reschedule the user in the current control interval.*

- *If feedback $F_1$ was received, schedule the other user.*

- *If feedback $F_2$ was received, calculate the expected immediate reward if the other user (identified as user $u$) is scheduled in the current interval (identified as $m$) as follows: $\pi_{m,u}\alpha$ where $\pi_{m,u}$ is the belief vector of user $u$ in the current control interval $m$. Now, schedule user $s$ is $\mathbf{p}_2\alpha \geq \pi_{m,u}\alpha$. Otherwise, schedule user $u$.*

*Proof.* Refer to Fig. 2. The argument for $F_3$ and $F_1$ are the same as in the previous case.

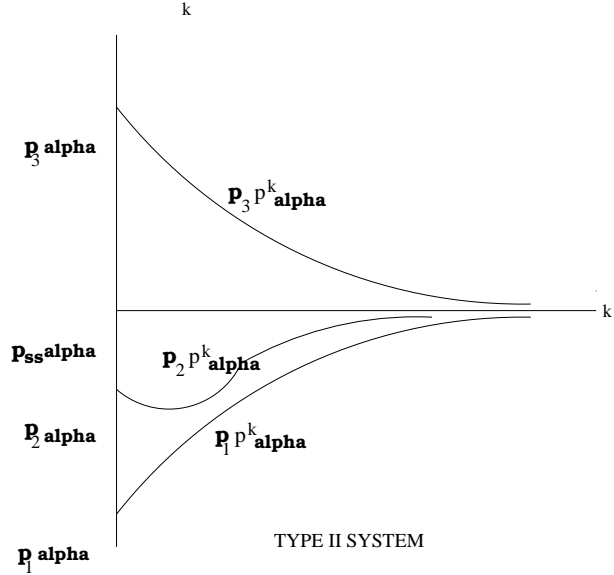

Figure 2: Type II system

When $F_2$ is received, as seen from the Fig. 2, the waiting user $u$ could have an expected reward greater than that of $s$ if $u$ had been dropped due to $F_1$ at least $k_0$ intervals earlier or if $\mathbf{p}_2 P^k \alpha$ does not monotonically increase to $\mathbf{p}_{ss}\alpha$ (Fig. 2 shows such a situation). Thus it is necessary to explicitly calculate the expected reward of user $u$ before making a greedy decision. □

Note that the results in Lemma 2-4 and hence the implementation structure of the greedy policy in Propositions 5-6 hold even when $\alpha_1 > 0$ as long as $\alpha_1 \leq \alpha_2 \leq \alpha_3$.

## 4 Comparison with the Genie aided system

In the two user, *two* state case, if the user scheduled (user $s$) in the previous control interval was observed to be in the best state, the scheduler retains the schedule (and hence accrues the best possible reward) since there is nothing more to gain by scheduling to the other user, while a loss is possible if the other user was in the worst state. Similarly, if user $s$ was observed to be in the worst state, the scheduler switches to the other, since there is nothing more to lose by scheduling to the other user (as compared to scheduling $s$ again), while a gain is possible if the other user was in the best state. Thus the two user, two state system is equivalent, in performance, to a genie-aided system where the scheduler learns about the states of both the users at the end of every interval.

This equivalence does not hold in the three state system. The *nothing more to gain* argument works when $s$ was observed to be in state 3 and the *nothing more to lose* argument works when $s$ was observed to be in state 1. However, when $s$ was observed to be in state 2, i.e., when $F_2$ was received, by scheduling to the other user (user $u$), the scheduler may either gain (if $u$ was in state 3) or lose (if $u$ was in state 3) as compared to when it schedules $s$ again. Thus with information about the state of the other user, there is definitely a room for improvement. Thus the three state (in general, more than two states) system is not equivalent to the genie-aided system. Note that, the genie aided system can be redefined as follows: the scheduler learns about the state of both the users iff $s$ was observed in state 2. We see from the discussion so far that this modified definition does not impart any performance loss in the genie-aided system.

From the preceding discussion, it can be seen that the original three user system approaches the genie-aided system under any of the following cases:

- $\mathbf{p}_2 \alpha = \mathbf{p}_3 \alpha$. Thus on receiving $F_2$ from user $s$, nothing more can be gained by scheduling the other user $u$ (while a loss is possible on switching). Hence, $s$ is rescheduled. Thus there is no need to learn the previous control interval state of $u$.

- $\mathbf{p}_2 \alpha = \mathbf{p}_1 \alpha$. Thus on receiving $F_2$ from user $s$, nothing more can be lost by scheduling the other user $u$ (while a gain is possible on switching). Hence, $u$ is scheduled. Again, there is no need to learn the previous control interval state of $u$.

With mathematical analysis, it can be seen that case 1 is possible iff $\alpha_2 = \alpha_3$ and $p_{21} = p_{31}$. While case 2 is possible iff $\alpha_2 = \alpha_3$ and $p_{11} = p_{21}$. When the first set of conditions is satisfied, it can be seen that the states 2 and 3 can be merged at a very generic level (not specific to the type of information used for scheduling) with the reduced transition matrix given as below:

$$\begin{bmatrix} p_{11} & p_{12} + p_{13} \\ p_{21} & p_{22} + p_{23} \end{bmatrix} \qquad (9)$$

where row 1 and column 1 corresponds to state 1 and row2 and column2 corresponds to the merged state. Thus the channel is effectively modeled by a two-state Markov chain thus explaining the equivalence with the genie-aided system.

However, it is interesting to note that, when the second set of conditions is satisfied, such a merger is not possible between states 1 and 2 since we still have $p_{13} \leq p_{23}$ making them different in their relationship with state 3. However, in the context of the ARQ based scheduling problem (specifically case 2 in the preceding discussion), they are synonymous and render the original system equivalent to the genie-aided system.

## 5   Bounds On the System Sum Reward Capacity

**Proposition 7.** *For the type I system, a lower bound to the sum reward capacity, $S_{LB,I}$, is given as*

$$S_{LB,I} \geq \mathbf{p}_2 \alpha - \mathbf{p}_{ss}^2(1)(\mathbf{p}_2 \alpha - \mathbf{p}_1 \alpha) \qquad (10)$$

*where $\mathbf{p}_{ss}^2(1)$ is the steady state probability that the state of the user is 1.*

This bound is obtained by replacing expected reward given $F_3$, i.e., $\mathbf{p}_3\alpha$ with $\mathbf{p}_2\alpha$ in the sum reward evaluation of the greedy policy. Thus this is in fact a lower bound to the greedy policy. Note that $S_{LB,I}$ decreases as the steady state probability of the less rewarding state 1 ($p_{ss}(1)$) increases. Also notice that as $\mathbf{p}_1\alpha \to \mathbf{p}_2\alpha$, $S_{LB} \to \mathbf{p}_2\alpha$. This is expected in light of the approach we used in obtaining $S_{LB}$, since the only reward that we accrue in any control interval is now $\mathbf{p}_2\alpha$. Also, the bound approaches the system sum reward capacity when states 2 and 3 become increasingly synonymous. This happens as $\alpha_2 \to \alpha_3$ and $p_{31} \to p_{21}$. The last statement comes from our discussion in the previous section, on the equivalence with the genie-aided system.

**Proposition 8.** *For the type II system, a lower bound to the sum reward capacity is given as*

$$S_{LB,II} = (2\mathbf{p}_{ss}(3) - \mathbf{p}_{ss}^2(3))\mathbf{p}_3\alpha + (1 - \mathbf{p}_{ss}(3))^2 \mathbf{p}_3\alpha \qquad (11)$$

The proof proceeds as follows: In any control interval the expected immediate reward after a feedback $F_2$ is received in the previous interval is replaced by the reward that would be expected if the other (not scheduled in the previous interval) user were scheduled. Note that, by the implementation structure of the greedy policy, this latter reward is $\leq$ the reward corresponding to the greedy choice[2]. Next we replace $\mathbf{p}_2\alpha$ with $\mathbf{p}_1\alpha$ giving the sum reward capacity lower bound.

Note that $S_{LB,II}$ is the same as a two user system that accrues reward $\mathbf{p}_3\alpha$ if at least one of the users are in state 3 and reward $\mathbf{p}_1\alpha$ if none of them are in state 3. This interpretation is strikingly similar to the interpretation we made in the two-state tow user problem in our preliminary research. However, note that the present interpretation does not yield to the case when the state of both users are available. For instance, if none of the users is in state 3 and at least one of them is in state 2, then, ideally, if the states of both the users are known, a reward of $\mathbf{p}_2\alpha$ must be accrued instead of $\mathbf{p}_1\alpha$. This demonstrates the loss in performance due to lack of knowledge of both user states, thus differentiating the 3-state system from the 2-state system.

**Proposition 9.** *An upper bound to the system sum reward capacity is given as*

$$S_{UB} = (2\mathbf{p}_{ss}(3) - \mathbf{p}_{ss}^2(3))\mathbf{p}_3\alpha + (2\mathbf{p}_{ss}(1)\mathbf{p}_{ss}(2) + \mathbf{p}_{ss}^2(2))\mathbf{p}_2\alpha + \mathbf{p}_{ss}^2(1)\mathbf{p}_1\alpha \qquad (12)$$

The bound is actually the sum reward capacity of the genie-aided system. Here if at least one of the users was in state 3 in the previous interval, the greedy policy schedules that user and accrues a reward $\mathbf{p}_3\alpha$. If none of the users were in state 3 but at least one of them in state 2, that user is scheduled and a reward of $\mathbf{p}_2\alpha$ is accrued. If both the users were in state 1, a reward of $\mathbf{p}_1\alpha$ is accrued.

# 6 On the Optimality of the Greedy Policy

We proceed by introducing the following properties of the $P$ matrix.

---

[2] The replacement is only with respect to the accrued reward in the sum reward expression, while the actual schedule decision is always maintained as greedy, so as not to disturb the initial conditions of the problem for the future intervals.

**Lemma 10.** *When $\mathbf{p}_2 P[001]^T \leq p_{23}$ (condition (A)), then $\mathbf{p}_2 P^{k+1}[001]^T \leq \mathbf{p}_2 P^k[001]^T$ $\forall k \geq 0$. Also the steady state element $p_{ss}(3) \leq p_{23}$ and $p_2 P^k[001]^T$ monotonically decreases to $p_{ss}(3)$ as $k \to \infty$. (A) is also a necessary condition for the preceding statement to hold.*

**Lemma 11.** *Under (A) from previous lemma, $\mathbf{p}_1 P^k[001]^T$ monotonically increases to $\mathbf{p}_{ss}(3)$ as $k \to \infty$, i.e, $\mathbf{p}_1 P^{k+1}[001]^T \geq \mathbf{p}_1 P^k[001]^T \forall k \in 0, 1, 2, \ldots$ and $\mathbf{p}_1 \lim_{k \to \infty} P^k[001]^T = \mathbf{p}_{ss}(3) \leq p_{23}$*

This result can be obtained by replacing $\alpha$ in Lemma 4 by $[0\ 0\ 1]^T$, and using $\mathbf{p}_{ss}(3) \leq p_{23}$ from Lemma 10.

**Proposition 12.** *When $p_{12} = p_{22} = p_{32}$ and $p_{23}p_{31} \geq p_{21}p_{13}$, greedy policy is optimal among the policies that retain the schedule when feedback $F_3$ is received.*

**Conjecture 13.** *When $p_{12} = p_{22} = p_{32}$ and $p_{23}p_{31} \geq p_{21}p_{13}$, greedy policy is globally optimal.*

The premise behind our conjecture is that, in light of the positive correlation property of the Markov chain, there is no obvious reason why the globally optimal policy would reject an user that was in the best state possible in the previous control interval. Thus it appears that the globally optimal policy belongs to the class that retains the schedule when $F_3$ is received suggesting that it is indeed the greedy policy itself.

# 7 Conclusion

We have considered the problem of scheduling under partial channel state information assumption in a Markov-modeled two-user downlink system with a channel state feedback provision. We classified the system in two types based on the transition probability matrix of the Markov chains and the reward structure. For each type, we established the implementation structure of the greedy policy. For the type I system, we showed that the greedy policy can be implemented via a simple round robin algorithm as was seen in our earlier work for the two-state Markov model. We studied the conditions under which the original system simplifies to the genie aided system and provided insights on these conditions. By appropriately bounding the immediate reward accrued in any control interval, we obtained bounds to the sum reward capacity of the system. Under some conditions on the $P$ matrix, by restricting the search space to a specific class of schedulers, we showed that the greedy policy is 'constrained search space' optimal and conjectured, with reasons, that the greedy policy is globally optimal as well.

# A Proof of Lemma 2

Let $\beta = [\beta_1 \ \beta_2 \ \beta_3]^T$, with $\beta_1 \leq \beta_2 \leq \beta_3$. Consider the inequality $\mathbf{p}_3 \beta \geq \mathbf{p}_2 \beta$. This can be rewritten as,

$$\begin{aligned}
\beta_1 p_{31} + \beta_2 p_{32} + \beta_3 p_{33} &\geq \beta_1 p_{21} + \beta_2 p_{22} + \beta_3 p_{23} \\
\Leftrightarrow \beta_1(p_{31} - p_{21}) &\geq \beta_2(p_{22} - p_{32}) + \beta_3(p_{23} - p_{33}) \\
\Leftrightarrow \beta_1(p_{21} - p_{31}) &\leq -\beta_2(p_{22} - p_{32}) + \beta_3(p_{33} - p_{23})
\end{aligned} \tag{13}$$

Since $\beta_2 \geq \beta_1$, it is now sufficient to prove $\beta_2(p_{21} - p_{31} + p_{22} - p_{32}) \leq \beta_3(p_{33} - p_{23})$, i.e., $\beta_2(p_{33} - p_{23}) \leq \beta_3(p_{33} - p_{23})$ which is indeed true. Consider the inequality $\mathbf{p}_2 \beta \geq \mathbf{p}_1 \beta$,

$$\begin{aligned}
\beta_1 p_{21} + \beta_2 p_{22} + \beta_3 p_{23} &\geq \beta_1 p_{11} + \beta_2 p_{12} + \beta_3 p_{13} \\
\Leftrightarrow \beta_2 + p_{23}(\beta_3 - \beta_2) - p_{21}(\beta_2 - \beta_1) &\geq \beta_2 + p_{13}(\beta_3 - \beta_2) - p_{11}(\beta_2 - \beta_1)
\end{aligned} \tag{14}$$

The last inequality is indeed true, since $p_{23} \geq p_{13}$, $p_{21} \leq p_{11}$ and $\beta_3 \geq \beta_2 \geq \beta_1$. Thus if $\beta_1 \leq \beta_2 \leq \beta_3$ and $\beta = [\beta_1 \ \beta_2 \ \beta_3]^T$,

$$\mathbf{p}_3 \beta \geq \mathbf{p}_2 \beta \geq \mathbf{p}_1 \beta \tag{15}$$

We can write, for $i \in 1, 2, 3$, $\mathbf{p}_i P^{k+1} \alpha = \mathbf{p}_i [\mathbf{p}_1 P^k \alpha \ \mathbf{p}_2 P^k \alpha \ \mathbf{p}_3 P^k \alpha]^T$. Thus if $\mathbf{p}_1 P^k \alpha \leq \mathbf{p}_2 P^k \alpha \leq \mathbf{p}_3 P^k \alpha$, we have, using (15), $\mathbf{p}_1 P^{k+1} \alpha \leq \mathbf{p}_2 P^{k+1} \alpha \leq \mathbf{p}_3 P^{k+1} \alpha$. Since $\alpha_1 = 0 \leq \alpha_2 \leq \alpha_3 = 1$, the lemma is established using induction.

# B Proof of Lemma 3 and Lemma 4

Consider $\mathbf{p}_3 P^{k+1} \alpha = p_{31} \mathbf{p}_1 P^k \alpha + p_{32} \mathbf{p}_2 P^k \alpha + p_{33} \mathbf{p}_3 P^k \alpha$. Since $\mathbf{p}_1 P^k \alpha \leq \mathbf{p}_2 P^k \alpha \leq \mathbf{p}_3 P^k \alpha$ from Lemma 2, we have $\mathbf{p}_3 P^{k+1} \alpha \leq \mathbf{p}_3 P^k \alpha$. Lemma 4 can be proved similarly.

# C Proof of Lemma 10

Let $\mathbf{p}_2 P^k [001]^T \leq \mathbf{p}_2 P^{k-1} [001]^T$. Multiplying both sides by $p_{22}$ and adding to both sides $p_{21} \mathbf{p}_1 P^{k-1} [001]^T + p_{23} \mathbf{p}_3 P^{k-1} [001]^T$,

$$p_{21} \mathbf{p}_1 P^{k-1}[001]^T + p_{22} \mathbf{p}_2 P^k[001]^T + p_{23} \mathbf{p}_3 P^{k-1}[001]^T \leq \mathbf{p}_2 P^k[001]^T \tag{16}$$

If we show that $p_{21} \mathbf{p}_1 P^k[001]^T + p_{23} \mathbf{p}_3 P^k[001]^T \leq p_{21} \mathbf{p}_1 P^{k-1}[001]^T + p_{23} \mathbf{p}_3 P^{k-1}[001]^T$, then using (16), $p_{21} \mathbf{p}_1 P^k[001]^T + p_{22} \mathbf{p}_2 P^k[001]^T + p_{23} \mathbf{p}_3 P^k[001]^T \leq \mathbf{p}_2 P^k[001]^T$, i.e, $\mathbf{p}_2 P^{k+1}[001]^T \leq \mathbf{p}_2 P^k[001]^T$. Consider the inequality

$$\begin{aligned}
p_{21}\mathbf{p}_1 P^k[001]^T + p_{23}\mathbf{p}_3 P^k[001]^T &\leq p_{21}\mathbf{p}_1 P^{k-1}[001]^T + p_{23}\mathbf{p}_3 P^{k-1}[001]^T \\
\Leftrightarrow \mathbf{p}_2 P^{k+1}[001]^T - p_{22}\mathbf{p}_2 P^k[001]^T &\leq \mathbf{p}_2 P^k[001]^T - p_{22}\mathbf{p}_2 P^{k-1}[001]^T \\
\Leftrightarrow \mathbf{p}_2(P^{k+1}[001]^T - P^k[001]^T) &\leq p_{22}(\mathbf{p}_2 P^k[001]^T - \mathbf{p}_2 P^{k-1}[001]^T) \\
\Rightarrow \mathbf{p}_2 P^{k+1}[001]^T &\leq \mathbf{p}_2 P^k[001]^T
\end{aligned} \tag{17}$$

where the last inequality is from the initial assumption that $\mathbf{p}_2 P^k[001]^T - \mathbf{p}_2 P^{k-1}[001]^T \leq 0$.

With $\mathbf{p}_2 P^1[001]^T \leq \mathbf{p}_2 P^0[001]^T$, i.e, $\mathbf{p}_2 P[001]^T \leq p_{23}$, using induction, we have the $\mathbf{p}_2 P^{k+1}[001]^T \leq \mathbf{p}_2 P^k[001]^T \ \forall k \geq 0$. Since steady state exists, by the definition of steady state, $\lim_{k\to\infty} P^k = \begin{bmatrix} \mathbf{p}_{ss} \\ \mathbf{p}_{ss} \\ \mathbf{p}_{ss} \end{bmatrix}$. Thus $\mathbf{p}_2 \lim_{k\to\infty} P^k[001]^T = \mathbf{p}_{ss}(3)$ and $\mathbf{p}_{ss}(3) \leq p_{23}$ by the monotonic decrease property of $\mathbf{p}_2 P^k[001]^T$. Also note that the direction of the inequalities throughout this proof can be changed and we can prove that $\mathbf{p}_2 P^k[001]^T$ monotonically *increases* to $\mathbf{p}_{ss}(3)$ as $k \to \infty$ if $\mathbf{p}_2 P[001]^T \geq p_{23}$. This establishes that $\mathbf{p}_2 P[001]^T \leq p_{23}$ is a necessary condition for the first part of the Lemma to hold.

## D  Proof of Proposition 12

Let the probability transition matrix satisfy the following conditions:

$$p_{12} = p_{22} = p_{32} \tag{18}$$

$$p_{23}p_{31} \geq p_{21}p_{13} \tag{19}$$

The preceding inequality along with condition (18) is equivalent to condition (A) in Lemma 10. Thus under (18) and (19), both Lemma 10 and Lemma 11 hold true. From Lemma 10, $p_{23} \geq \mathbf{p}_{ss}(3)$. From (18), $p_{ss}(2) = p_{22}$. Thus $p_2\alpha - p_{ss}\alpha = p_{22}\alpha_2 + p_{23} - \mathbf{p}_{ss}(2)\alpha_2 - \mathbf{p}_{ss}(3) = p_{23} - \mathbf{p}_{ss}(3) \geq 0$. The system is thus type I.

Consider a control interval $m > 1$ with belief vectors $\pi_{m,1}$, $\pi_{m,2}$ and action $a_m$. If we can show for any $m$ that, assuming the greedy policy will be implemented in all the future control intervals, the greedy policy is optimal in control interval $m$, then using induction from interval 1, where greedy is indeed optimal, we could establish the long term optimality of the greedy policy. Let $\{\mathfrak{A}_k\}_{k\leq m-1} = \{\widehat{\mathfrak{A}}_k\}_{k\leq m-1}$ and let $S_k$ be the state vector such that $S_k(i)$ is the state of the channel of user $i$ in interval $k$. We rewrite the net expected reward as follows

$$V_m(\pi_{m,1}, \pi_{m,2}, \{a_m, \{\widehat{\mathfrak{A}}_k\}_{k\leq m-1}\}) = \pi_{m,a_m}\alpha + \sum_{S_m} P_{S_m|\pi_{m,1},\pi_{m,2}}(S_m|\pi_{m,1},\pi_{m,2})\hat{V}_{m-1}(S_m, \hat{a}_{m-1}),$$

where $\hat{V}_{m-1}$ is the expected future reward conditioned on the state vector in control interval $m$. The *hat* on this quantity emphasizes the use of the greedy policy in all $k \leq m - 1$. $P_{S_m|\pi_{m,1},\pi_{m,2}}(S_m|\pi_{m,1},\pi_{m,2})$ is the conditional probability of the current state vector $S_m$ given the belief vectors $\pi_{m,1}$, $\pi_{m,2}$. The scheduling decision in the next control interval, $\hat{a}_{m-1}$, is based on the greedy policy and is a function of the ARQ feedback received in the current control interval $k$, i.e., $S_m(a_m)$. The decision logic was summarized in Proposition 5. We now proceed to compare the net expected reward when $a_m = 1$ and $a_m = 2$. The net expected reward when $a_m = 1$ is written as follows,

$$\begin{aligned}
&V_m(\pi_{m,1}, \pi_{m,2}, \{a_m = 1, \{\widehat{\mathfrak{A}}_k\}_{k \leq m-1}\}) \\
&= \pi_{m,1}\alpha + P_{S_m|\pi_{m,1},\pi_{m,2}}(S_m = [1\ 1]|\pi_{m,1}, \pi_{m,2})\hat{V}_{m-1}(S_m = [1\ 1], \hat{a}_{m-1} = 2) \\
&\quad + P_{S_m|\pi_{m,1},\pi_{m,2}}(S_m = [1\ 2]|\pi_{m,1}, \pi_{m,2})\hat{V}_{m-1}(S_m = [1\ 2], \hat{a}_{m-1} = 2) \\
&\quad + P_{S_m|\pi_{m,1},\pi_{m,2}}(S_m = [1\ 3]|\pi_{m,1}, \pi_{m,2})\hat{V}_{m-1}(S_m = [1\ 3], \hat{a}_{m-1} = 2) \\
&\quad + P_{S_m|\pi_{m,1},\pi_{m,2}}(S_m = [2\ 1]|\pi_{m,1}, \pi_{m,2})\hat{V}_{m-1}(S_m = [2\ 1], \hat{a}_{m-1} = 1) \\
&\quad + P_{S_m|\pi_{m,1},\pi_{m,2}}(S_m = [2\ 2]|\pi_{m,1}, \pi_{m,2})\hat{V}_{m-1}(S_m = [2\ 2], \hat{a}_{m-1} = 1) \\
&\quad + P_{S_m|\pi_{m,1},\pi_{m,2}}(S_m = [2\ 3]|\pi_{m,1}, \pi_{m,2})\hat{V}_{m-1}(S_m = [2\ 3], \hat{a}_{m-1} = 1) \\
&\quad + P_{S_m|\pi_{m,1},\pi_{m,2}}(S_m = [3\ 1]|\pi_{m,1}, \pi_{m,2})\hat{V}_{m-1}(S_m = [3\ 1], \hat{a}_{m-1} = 1) \\
&\quad + P_{S_m|\pi_{m,1},\pi_{m,2}}(S_m = [3\ 2]|\pi_{m,1}, \pi_{m,2})\hat{V}_{m-1}(S_m = [3\ 2], \hat{a}_{m-1} = 1) \\
&\quad + P_{S_m|\pi_{m,1},\pi_{m,2}}(S_m = [3\ 3]|\pi_{m,1}, \pi_{m,2})\hat{V}_{m-1}(S_m = [3\ 3], \hat{a}_{m-1} = 1) \quad (20)
\end{aligned}$$

Note that the scheduler uses the information of the state of the scheduled user (user 1) alone in the scheduling decisions, consistent with the problem setup. Also note that when $S_m(1) = 2$, the schedule is retained. This is consistent with the implementation structure of the greedy policy seen in Proposition 5, where the scheduler *retains the scheduling choice even $F_2$ is received*. As was discussed in the same proposition, this is a greedy decision only if an user was never dropped in the past for giving feedback $F_3$. Since we are restricting to the class of schedulers that retains the schedule when $F_3$ is satisfied[3], this is indeed a greedy decision. Since the Markov channel statistics are identical across the users, we have $\hat{V}_k(S_{k+1} = [x\ y], \hat{a}_k = 1]) = \hat{V}_k(S_{k+1} = [y\ x], \hat{a}_k = 2])$. Expanding the net expected reward when $a_m = 2$ along the lines of (20) and using the preceding symmetry property, we have,

$$\begin{aligned}
&V_m(\pi_{m,1}, \pi_{m,2}, \{a_m = 1, \{\widehat{\mathfrak{A}}_k\}_{k \leq m-1}\}) - V_m(\pi_{m,1}, \pi_{m,2}, \{a_m = 2, \{\widehat{\mathfrak{A}}_k\}_{k \leq m-1}\}) \\
&= \pi_{m,1}\alpha - \pi_{m,2}\alpha \\
&\quad + \left[\hat{V}_{m-1}(S_m = [3\ 2], \hat{a}_{m-1} = 1) - \hat{V}_{m-1}(S_m = [2\ 3], \hat{a}_{m-1} = 1)\right] \times \\
&\quad \left[\pi_{m,1}(3)\pi_{m,2}(2) - \pi_{m,1}(2)\pi_{m,2}(3)\right] \quad (21)
\end{aligned}$$

Let $\hat{a}_m$ indicate the greedy choice among the users in the current control interval, i.e., $\hat{a}_m = \arg\max_{i \in 1,2} R_m(\pi_{m,i})$. Let $\tilde{a}_m$ indicate the other user. The net expected reward can now be rewritten as,

$$\begin{aligned}
&V_m(\pi_{m,1}, \pi_{m,2}, \{a_m = \hat{a}_m, \{\widehat{\mathfrak{A}}_k\}_{k \leq m-1}\}) - V_m(\pi_{m,1}, \pi_{m,2}, \{a_m = \tilde{a}_m, \{\widehat{\mathfrak{A}}_k\}_{k \leq m-1}\}) \\
&= \pi_{m,\hat{a}_m}\alpha - \pi_{m,\tilde{a}_m}\alpha \\
&\quad + \left[\hat{V}_{m-1}(S_m = [3\ 2], \hat{a}_{m-1} = 1) - \hat{V}_{m-1}(S_m = [2\ 3], \hat{a}_{m-1} = 1)\right] \times \\
&\quad \left[\pi_{m,\hat{a}_m}(3)\pi_{m,\tilde{a}_m}(2) - \pi_{m,\hat{a}_m}(2)\pi_{m,\tilde{a}_m}(3)\right] \quad (22)
\end{aligned}$$

---

[3]This is the only instance in the proof where we constrain the search space.

where, by definition, $\pi_{m,\hat{a}_m}\alpha \geq \pi_{m,\tilde{a}_m}\alpha$. We now proceed to show that the quantity $\hat{V}_{m-1}(S_m = [3\ 2], \hat{a}_{m-1} = 1) - \hat{V}_{m-1}(S_m = [2\ 3], \hat{a}_{m-1} = 1)$ is non-negative. With $\hat{V}_k(S_{k+1} = [x\ y]) := \hat{V}_k(S_{k+1} = [x\ y], \hat{a}_k = 1)$, and expanding $\hat{V}_{m-1}(S_m = [x\ y])$ along the lines of (20) with $\pi_{m-1,1} = \mathbf{p}_x$ and $\pi_{m-1,2} = \mathbf{p}_y$ and $a_{m-1} = 1$, we have the following.

$$\hat{V}_{m-1}(S_m = [3\ 2]) - \hat{V}_{m-1}(S_m = [2\ 3])$$
$$= \mathbf{p}_3\alpha - \mathbf{p}_2\alpha + \left[\hat{V}_{m-2}(S_{m-1} = [3\ 2]) - \hat{V}_{m-2}(S_{m-1} = [2\ 3])\right](p_{33}p_{22} - p_{23}p_{32}) \quad (23)$$

By the property of the $P$ matrix, $p_{33} \geq p_{23}$ and $p_{22} \geq p_{32}$. Also, we have seen in Lemma3 that $r_3 \geq r_2 \geq r_1$. Thus if $\hat{V}_{m-2}(S_{m-1} = [3\ 2]) - \hat{V}_{m-2}(S_{m-1} = [2\ 3]) \geq 0$, then $\hat{V}_{m-1}(S_m = [3\ 2]) - \hat{V}_{m-1}(S_m = [2\ 3]) \geq 0$. Expanding $\hat{V}_{m-2}(S_{m-1} = [3\ 2]) - \hat{V}_{m-2}(S_{m-1} = [2\ 3]) \geq 0$ along the lines of (23) repeatedly and using $\hat{V}_1(S_m = [3\ 2]) - \hat{V}_1(S_m = [2\ 3]) = r_3 - r_2 \geq 0$, by induction, we can show that $\hat{V}_{m-2}(S_{m-1} = [3\ 2]) - \hat{V}_{m-2}(S_{m-1} = [2\ 3]) \geq 0$. Thus $\hat{V}_{m-1}(S_m = [3\ 2]) - \hat{V}_{m-1}(S_m = [2\ 3]) \geq 0$. Applying this inequality in (22), we see that the optimality of the greedy policy (in the specified class of policies) can be established if we show that the following condition (condition (S)) holds:

$$\pi_{m,\hat{a}_m}(3)\pi_{m,\tilde{a}_m}(2) \geq \pi_{m,\hat{a}_m}(2)\pi_{m,\tilde{a}_m}(3). \quad (24)$$

It appears that the preceding condition is too generic to hold true. However, by constraining the belief vectors to the set of values that will be encountered in the ARQ based scheduling problem, we will now show that, (24) indeed holds true.

We first introduce the following result: From Lemma 10, $\mathbf{p}_2 P^k [001]^T$ monotonically decreases to $\mathbf{p}_{ss}[001]^T = \mathbf{p}_{ss}(3)$ as $k$ increases. Since $\mathbf{p}_2 P^k[010] = p_{22} = \mathbf{p}_{ss}(2)$, the expected reward from an user given the channel of the user was in state 2 $k+1$ intervals earlier, given by, $\mathbf{p}_2 P^k \alpha = \alpha_{(2)}\mathbf{p}_{ss}(2) + \mathbf{p}_2 P^k[001]^T$ monotonically decreases to $\mathbf{p}_{ss}\alpha$.

We proceed with studying the sufficient condition under various belief vectors encountered in the ARQ based scheduling problem. Assume the scheduling process has begun in a control interval earlier than $m$ and is performed uninterrupted till the horizon, i.e, control interval 1 - assumption (A)[4]. The belief vector of the greedy choice $\hat{a}_m$ and the other user $\tilde{a}_m$, for the type I system under consideration, falls under one of the following cases.

- 1. User $\hat{a}_m$ was scheduled in the previous control interval, $m+1$, and had given a feedback $F_3$. The belief vector $\pi_{m,\hat{a}_m} = \mathbf{p}_3$. The other user was either scheduled in $k+1$ control intervals earlier (with $k \in 1, 2, \ldots$) with any of the three possible feedback or was never scheduled in the past. Thus the belief vector of $\tilde{a}_m$ is of the form $\mathbf{p}_i P^k$ with $i \in 1, 2, 3$ and $k \in 1, 2, \ldots$. Note that if $\tilde{a}_m$ was never scheduled in the past, then $\pi_{m,\tilde{a}_m} = \mathbf{p}_{ss}$ which still falls in the preceding form.

---
[4]Note that there is no loss of generality in this assumption for the following reason: The problem setup and the optimality analysis of any policy implicitly assumes uninterrupted scheduling until the horizon. This is to be in tune with the interval to interval evolution of the underlying Markov chains. Thus when the uninterrupted scheduling process begins at a control interval $M$, for all $m < M$ condition (A) is satisfied automatically. In the control interval $M$, however, part of the condition, i.e *scheduling process began earlier*, does not hold. But at the origin, i.e., the control interval $M$, the belief vectors of all the users take the steady state value, $\mathbf{p}_{ss}$. Thus, by all symmetry, the question of what scheduling decision to make and hence the question of the optimality of the greedy policy at $M$ becomes irrelevant.

- 2. User $\tilde{a}_m$ was scheduled in the previous control interval and had given a feedback $F_1$. User $\hat{a}_m$ was either scheduled $k+1$ control intervals earlier (with $k \in 1, 2, \ldots$) with any of the three possible feedbacks or was never scheduled in the past. The belief vectors are given by $\pi_{m,\tilde{a}_m} = \mathbf{p}_1$ and $\pi_{m,\hat{a}_m} = \mathbf{p}_i P^k$ with $i \in 1, 2, 3$ and $k \in 1, 2, \ldots$.

- 3. User $\hat{a}_m$ was scheduled in the previous control interval and had given a feedback $F_2$. User $\tilde{a}_m$ was scheduled $k+1$ control intervals earlier (with $k \in 1, 2, \ldots$) with feedback $F_1$ or was never scheduled in the past. The belief vectors are given by $\pi_{m,\hat{a}_m} = \mathbf{p}_2$ $\pi_{m,\tilde{a}_m} = \mathbf{p}_1 P^k$ with $k \in 1, 2, \ldots$.

- 4. User $\hat{a}_m$ was scheduled in the previous control interval and had given a feedback $F_2$. User $\tilde{a}_m$ was scheduled $k+1$ control intervals earlier (with $k \in 1, 2, \ldots$) with feedback $F_2$. The belief vectors are given by $\pi_{m,\hat{a}_m} = \mathbf{p}_2$ $\pi_{m,\tilde{a}_m} = \mathbf{p}_2 P^k$ with $k \in 1, 2, \ldots$.

- 5. User $\hat{a}_m$ was scheduled in the previous control interval and had given a feedback $F_2$. User $\tilde{a}_m$ was scheduled $L+1$ or more control intervals earlier with feedback $F_3$. $L$ is the number of coherence intervals such that, reward expected from an user that was observed to be in state 2 in the previous control interval is higher than the reward expected from an user that was observed in state 3 $k+1$ control intervals earlier iff $k \geq L$. Mathematically, $L$ is such that,

$$\mathbf{p}_2 \alpha \geq \mathbf{p}_3 P^k \alpha \; if \; k \geq L \mathbf{p}_2 \alpha < \mathbf{p}_3 P^k \alpha \; if \; k < L \qquad (25)$$

Note that such an $L$ exists since $\mathbf{p}_2 \alpha \leq \mathbf{p}_3 \alpha$ and both $\mathbf{p}_2 P^k \alpha$ and $\mathbf{p}_3 P^k \alpha$ monotonically decreases (with $k$) to $\mathbf{p}_{ss} \alpha \leq \mathbf{p}_2 \alpha$. The belief vectors are hence given as $\pi_{m,\hat{a}_m} = \mathbf{p}_2$, $\pi_{m,\tilde{a}_m} = \mathbf{p}_3 P^k$ with $k \geq L$.

- 6. User $\tilde{a}_m$ was scheduled in the previous control interval and had given a feedback $F_2$. User $\hat{a}_m$ was scheduled $k+1$ control intervals earlier with feedback $F_3$ with $k < L$. The belief vectors are as follows: $\pi_{m,\hat{a}_m} = \mathbf{p}_3 P^k$ with $k < L$ and $\pi_{m,\tilde{a}_m} = \mathbf{p}_2$.

The above list is exhaustive. In fact, cases 5 and 6 will never appear since we are considering the class of schedulers that never drop an user when it sends an $F_3$. However, we will show that even for these cases the sufficient condition is satisfied. In all the above 6 cases, $R_m(\hat{a}_m) \geq R_m(\tilde{a}_m)$ as required by the definition of $\hat{a}_m$. We now focus on the sufficient condition (S) for each of the above cases.

- 1. Sufficient condition (S) is given as follows:

$$\begin{aligned} \pi_{m,\hat{a}_m}(3)\pi_{m,\tilde{a}_m}(2) &\geq \pi_{m,\hat{a}_m}(2)\pi_{m,\tilde{a}_m}(3) \\ i.e., \; p_{33}\mathbf{p}_i P^k [010]^T &\geq p_{32}\mathbf{p}_i P^k [001]^T, \forall i \in 1,2,3, \; k \in 1,2,\ldots \end{aligned} \qquad (26)$$

Since $p_{12} = p_{22} = p_{32}$, we have

$$\mathbf{p}_i P^k [010]^T = p_{12} = p_{22} = p_{32} \forall i \in 1,2,3, \ k \in 1,2,\ldots \quad (27)$$

Also, $\mathbf{p}_i P^k [001]^T = \mathbf{p}_i P^{k-1} P [001]^T = \mathbf{p}_i P^{k-1} [p_{13}\ p_{23}\ p_{33}]^T \leq p_{33}$, since $p_{33} \geq p_{23} \geq p_{13}$ by the property of the $P$ matrix. Thus (S) holds for case 1.

- 2. (S) is as follows: $\mathbf{p}_i P^k [001]^T p_{12} \geq \mathbf{p}_i P^k [010]^T p_{13}, \forall i \in 1,2,3, \ k \in 1,2,\ldots$.

  From the symmetry property (27), $p_{12} = \mathbf{p}_i P^k [010]^T$. Also since $p_{13} \leq p_{23} \leq p_{33}$ we can show $\mathbf{p}_i P^k [001]^T \geq p_{13}$. Thus (S) is satisfied for case 2.

- 3. (S): $p_{23} \mathbf{p}_1 P^k [010]^T \geq p_{22} \mathbf{p}_1 P^k [001]^T$. From Lemma 11, $p_1 P^k [001]^T$ monotonically increases to $p_s s(3)$ as $k$ increases as $0,1,2,\ldots$. Since $p_{23} \geq \mathbf{p}_{ss}(3)$ (using Lemma9), we have $p_{23} \geq \mathbf{p}_1 P^k [001]^T$. Also, $\mathbf{p}_1 P^k [010]^T = p_{22}$ from the symmetry property in (27). Thus (S) holds for case 3.

- 4. (S): $p_{23} \mathbf{p}_2 P^k [010]^T \geq p_{22} \mathbf{p}_2 P^k [001]^T$. From Lemma 10, $\mathbf{p}_2 P^k [001]^T$ monotonically decreases from $p_{23}$ to $\mathbf{p}_{ss}(3)$ as $k$ increases as $0,1,2,\ldots$. Thus $p_{23} \geq \mathbf{p}_2 P^k [001]^T$. This inequality along with the symmetry property (27) establishes (S) for case 4.

- 5. (S): $p_{23} \mathbf{p}_3 P^k [010]^T \geq p_{22} \mathbf{p}_3 P^k [001]^T$ with $k \geq L$. Note that for all $k \geq L$,

$$\begin{aligned} \mathbf{p}_2 \alpha &\geq \mathbf{p}_3 P^k \alpha \\ \Rightarrow \alpha_2 p_{22} + p_{23} &\geq \alpha_2 \mathbf{p}_3 P^k [010]^T + \mathbf{p}_3 P^k [001]^T \\ \Rightarrow p_{23} &\geq \mathbf{p}_3 P^k [001]^T \end{aligned} \quad (28)$$

  where we have used the symmetry property $p_{22} = \mathbf{p}_3 P^k [010]^T$ in obtaining the last inequality. (S) is established by using the symmetry property along with the preceding inequality.

- 6. (S): $\mathbf{p}_3 P^k [001]^T p_{22} \geq \mathbf{p}_3 P^k [010]^T p_{23}$ with $k < L$. For $k < L$, $\mathbf{p}_2 \alpha < \mathbf{p}_3 P^k \alpha$. Expanding both the sides along the lines of case 5 and using the symmetry property of (27), (S) can be established for case 6.

Thus the sufficient condition for the constrained search space optimality of the greedy policy is satisfied.

[16] S. H. A. Ahmad, M. Liu, T. Javadi, Q. Zhao, and B. Krishnamachari, "Optimality of mypoic sensing in multi-channel opportunistic access," *IEEE Trans. on Information Theory,* sub- mitted May 2008. (http://www.ece.ucdavis.edu/∼qzhao/Journal.html).

[17] Q. Zhao and B. Krishnamachari, "Optimality of mypoic polcty in opportunistic access with noisy observations," *IEEE Trans. on Automatic Control*, submitted Feb. 2008. (http://www.ece.ucdavis.edu/∼qzhao/Journal.html).

[18] S. Murugesan, P. Schniter, and N. B. Shroff, "Multiuser Scheduling in a Markov-modeled Downlink Environment," *Proc. Allerton Conf. on Communication, Control, and Computing,* (Monticello, IL), Sept. 2008.

[19] Marius Iosifescu, "Finite Markov Processes and Their Applications," *John Wiley and Sons*, 1980.

[20] R. D. Smallwood and E. J. Sondik, "The Optimal Control of Partially Observable Markov Processes Over a Finite Horizon," *Operations Research,* Sep. 1973.

[21] S. Christian Albright, "Structural Results for Partially Observable Markov Decision Processes," *Operations Research,* vol. 27, no. 5, pp. 1041-1053, Sep.-Oct. 1979.

[22] C. C. White and W. Scherer, "Solution procedures for partially observed Markov decision processes," *Operations Research,* pp. 791797, 1985.

[23] G. E. Monahan, "A survey of partially observable Markov decision processes: Theory, Models, and Algorithms," *Management Science,* vol. 28, no. 1, pp. 116, Jan. 1982.